\pdfoutput=1

\documentclass[11pt,twoside,a4paper,cmspaper,final,collab]{cms-tdr}
\def\svnVersion{340453}\def\cmsCernNoTag{CERN-PH-EP/2015-271}\def\cmsCernDate{\today}\def\cmsMessage{Published in Physical Review Letters as \href{http://dx.doi.org/10.1103/PhysRevLett.116.172302}{\doi{10.1103/PhysRevLett.116.172302.}}}

\begin{document}\cmsNoteHeader{FSQ-15-002}

\hyphenation{had-ron-i-za-tion}
\hyphenation{cal-or-i-me-ter}
\hyphenation{de-vices}
\RCS$Revision: 335541 $
\RCS$HeadURL: svn+ssh://svn.cern.ch/reps/tdr2/papers/FSQ-15-002/trunk/FSQ-15-002.tex $
\RCS$Id: FSQ-15-002.tex 335541 2016-03-24 11:05:59Z mguilbau $
\newlength\cmsFigWidth
\ifthenelse{\boolean{cms@external}}{\setlength\cmsFigWidth{0.98\columnwidth}}{\setlength\cmsFigWidth{0.7\textwidth}}
\ifthenelse{\boolean{cms@external}}{\providecommand{\cmsLeft}{top\xspace}}{\providecommand{\cmsLeft}{left\xspace}}
\ifthenelse{\boolean{cms@external}}{\providecommand{\cmsRight}{bottom\xspace}}{\providecommand{\cmsRight}{right\xspace}}
\newcommand{\Peight}{{\textsc{pythia8}}\xspace}
\newcommand{\roots}{\ensuremath{\sqrt{s}}}
\newcommand{\rootsNN}{\ensuremath{\sqrt{s_{_\mathrm{NN}}}}}
\newcommand{\deta}{\ensuremath{\Delta\eta}}
\newcommand{\dphi}{\ensuremath{\Delta\phi}}
\newcommand{\pp}{\ensuremath{\Pp\Pp}\xspace}
\newcommand{\pPb}{\ensuremath{\Pp\mathrm{Pb}}\xspace}
\newcommand{\PbPb}{\ensuremath{\mathrm{PbPb}}\xspace}
\providecommand{\EPOS}{\textsc{epos}\xspace}
\newcommand{\noff}{\ensuremath{N_\text{trk}^\text{offline}}\xspace}
\cmsNoteHeader{FSQ-15-002}
\title{Measurement of long-range near-side two-particle angular correlations in
\texorpdfstring{$\Pp\Pp$ collisions at $\sqrt{s} = 13$\TeV}{pp collisions at sqrt(s) = 13 TeV}}

\date{\today}

\abstract{
Results on two-particle angular correlations for charged particles
produced in $\Pp\Pp$ collisions at a center-of-mass energy of
13\TeV are presented. The data
were taken with the CMS detector at the LHC and correspond to an
integrated luminosity of about 270\nbinv. The correlations are studied
over a broad range of pseudorapidity ($\abs{\eta}<2.4$) and over the full
azimuth ($\phi$) as a function of charged particle multiplicity and
transverse momentum (\pt). In
high-multiplicity events, a long-range ($\abs{\Delta\eta}>2.0$),
near-side ($\Delta\phi\approx 0$) structure emerges in the
two-particle $\Delta\eta$--$\Delta\phi$ correlation
functions. The magnitude of the correlation exhibits a pronounced maximum in the
range $1.0 < \pt < 2.0$\GeVc and an approximately linear increase with the
charged particle multiplicity, with an overall correlation strength
similar to that found in earlier pp data at $\sqrt{s} = 7$\TeV. The present measurement
extends the study of near-side long-range correlations up to charged particle multiplicities
$N_\mathrm{ch} \sim 180$, a region so far unexplored in \pp collisions.
The observed long-range correlations are
compared to those seen in \pp, \pPb, and \PbPb collisions at lower collision energies.
}

\hypersetup{%
pdfauthor={CMS Collaboration},%
pdftitle={Measurement of long-range near-side two-particle angular correlations in pp collisions at sqrt(s) = 13 TeV},%
pdfsubject={CMS},%
pdfkeywords={CMS, physics, correlations, ridge, high-multiplicity, heavy-ion}}

\maketitle

Studies of particle correlations in high-energy hadron-hadron collisions provide valuable
information on the underlying quantum chromodynamics processes leading to particle production.
Measurements of two-particle angular correlations are typically performed in terms of
two-dimensional $\deta$--$\dphi$ correlation functions, where $\eta$ is the pseudorapidity and $\phi$ is the azimuthal angle.
Of particular interest in studies of possible novel partonic collective effects
is the long-range (e.g., $\abs{\Delta\eta} > 2.0$) structure
of two-particle correlation functions, in which the effects of known sources such as resonance decays and fragmentation of
high-momentum partons are known to be small. In most Monte Carlo (MC) event generators for proton-proton (\pp) collisions,
the typical sources of such long-range correlations are momentum conservation and
away-side ($\dphi \approx \pi$) jet correlations.
Measurements in high-energy nucleus-nucleus collisions have shown a long-range structure in the two-particle angular correlations
functions, which has been attributed to the presence of the hot and dense matter formed~\cite{Alver:2008gk}.
Several novel features were observed in azimuthal correlations
over large $\Delta\eta$ for intermediate particle transverse momenta,
$\pt \approx 1$--5\GeVc~\cite{Alver:2009id,Abelev:2009jv}. These correlations
are thought to arise from the response of a hydrodynamically expanding
partonic medium to fluctuations of the initial collision geometry
~\cite{Alver:2010gr,Alver:2010dn,Schenke:2010rr,Petersen:2010cw,Xu:2010du,Teaney:2010vd}.
Measurements in \pp collisions at a center-of-mass energy of $\roots = 7$\TeV have also revealed
the presence of long-range, near-side ($\dphi \approx 0$) correlations in events with
very large final-state particle multiplicity~\cite{Khachatryan:2010gv}. Similar phenomena have
also been observed in high-multiplicity proton-lead (\pPb)
collisions~\cite{CMS:2012qk,Abelev:2012ola,Aad:2012gla}, where they have been studied
extensively~\cite{Chatrchyan:2013nka,Khachatryan:2014jra,Khachatryan:2015waa,Khachatryan:2015oea,Aad:2013fja,Aad:2014lta,Abelev:2014mda,ABELEV:2013wsa}.

A wide range of models have been suggested to
explain the emergence of these correlations in \pp~\cite{Li:2012hc}
and \pPb~\cite{Bozek:2011if,Bozek:2012gr,Dusling:2012cg,Dumitru:2014vka,Schenke:2015aqa} collisions.
While models based on a hydrodynamic approach can describe many aspects of the observed correlations~\cite{Bozek:2011if,Bozek:2012gr},
it has been proposed that initial-state correlations of gluon fields
could also lead to similar effects~\cite{Dusling:2012cg,Dumitru:2014vka,Schenke:2015aqa}.

The LHC at CERN has recently started to deliver \pp collisions at a new energy regime at $\roots = 13$\TeV,
and there is renewed interest in investigating this phenomenon, especially its energy dependence.
The first measurement of long-range two-particle correlations
in \pp collisions at $\roots = 13$\TeV has been reported by the ATLAS collaboration~\cite{Aad:2015gqa}.
In this Letter, studies of long-range correlations in \pp collisions at $\roots = 13$\TeV 
with the CMS detector are presented. The measurements are performed over a wide range in
charged particle multiplicity and \pt. The strength of long-range
near-side correlations is quantified, and results for \pp, \pPb, and \PbPb systems
at various collision energies are compared.

The central feature of the CMS apparatus is a superconducting solenoid
of 6\unit{m} internal diameter, providing a magnetic field of 3.8\unit{T}. Within the solenoid
volume are a silicon pixel and strip tracker, a lead tungstate crystal
electromagnetic calorimeter (ECAL, $\abs{\eta}< 3$), and a brass and scintillator hadron
calorimeter (HCAL, $\abs{\eta}< 3$), each composed of a barrel and two endcap sections.
Extensive forward calorimetry (HF, $3 < \abs{\eta}< 5$) complements the coverage provided by the barrel and endcap detectors.
Muons are measured in gas-ionization detectors embedded in the steel
flux-return yoke outside the solenoid.
The silicon tracker measures charged particles within the pseudorapidity
range $\abs{\eta}< 2.5$. It consists of 1440 silicon pixel and 15\,148
silicon strip detector modules. For non-isolated particles of $1 < \pt < 10\GeVc$
and $\abs{\eta} < 1.4$, the track resolutions are typically 1.5\% in \pt
and 25--90 (45--150)\mum in the transverse (longitudinal) impact parameter~\cite{TRK-11-001}.
The first level (L1) of the CMS trigger system,
composed of custom hardware processors, uses information from the calorimeters and muon detectors
to select the most interesting events in a fixed time interval of less than 4\mus. The high-level
trigger (HLT) processor farm further decreases the event rate from around 100\unit{kHz} to less
than 1\unit{kHz}, before data storage.
A more detailed description of the CMS detector, together with a definition
of the coordinate system used and the relevant kinematic variables,
can be found in Ref.~\cite{Chatrchyan:2008zzk}.
The MC simulation of the CMS detector response is based on \GEANTfour~\cite{GEANT4}.

The data used in this study were recorded under special running conditions
in which the beams were separated at the CMS interaction point, resulting
in an average of 1.3 \pp interactions per bunch crossing.
The integrated luminosity recorded was about 270\nbinv.
As the average number of \pp interactions per bunch crossing is small in the present data,
minimum bias (MB) \pp events were selected online by simply requiring that two proton bunches
collide near the center of the CMS detector. Only a small fraction (${\sim}10^{-3}$) of all MB pp events
were recorded (i.e., the trigger was prescaled).
In order to enhance the fraction of high-multiplicity events, additional samples were collected with a dedicated
selection procedure that combined the CMS L1 and HLT systems.
At L1, the total transverse energy summed over ECAL and HCAL was required to be greater than a given threshold
(both 15 and 40\GeV thresholds were used). Only the lowest threshold trigger was prescaled. Track reconstruction for the HLT was based on
the three layers of the pixel detectors, and required that the track originates within a cylindrical region centered on the nominal interaction point. This region
has a length of 30~cm along the beam direction and a radius of 0.2\unit{cm} perpendicular to it.
For each event, the vertex reconstructed with the highest number of tracks was selected.
The number of tracks (${N}_\text{trk}^\text{online}$)
with $\abs{\eta}<2.4$, $\pt > 0.4\GeVc$, and a distance of closest approach of 0.12\unit{cm} or
less from this vertex was determined for each event. Data were taken with thresholds
${N}_\text{trk}^\text{online}>$60 or 85 (based on events selected with a L1 total energy larger than 15\GeV), and $110$
(based on events selected with a L1 total energy larger than 40\GeV).

In the offline analysis, hadronic collisions are selected by requiring at least one tower
in each of the two HF calorimeters with more than 3\GeV energy
to suppress diffractive interactions~\cite{Khachatryan:2010xs}.
Events are also required to contain at least one reconstructed primary vertex
with a position along the beam axis, $z_\text{vtx}$, within 15\unit{cm} of the nominal interaction point and within 0.15\unit{cm}
of the beams in the transverse plane. In addition, at least two tracks must be associated to this vertex.
As the data have an average of 1.3 \pp interactions per bunch crossing,
a substantial fraction of events have at least one additional interaction (pile-up).
A procedure similar to that described in Ref.~\cite{Chatrchyan:2013nka}
is used for identifying and rejecting pile-up events. It is based on the number of tracks associated with each
reconstructed vertex and the distance between multiple vertices.
If the distance between the highest-multiplicity vertex
and the closest additional vertex along the $z$ direction is larger than 1\unit{cm}, the event is accepted.
This is because the tracks used for the correlation analysis
are always selected with respect to the highest-multiplicity vertex in the event.
An additional vertex sufficiently far from the highest-multiplicity
vertex has a negligible effect on the analysis.
The MC studies carried out with the \EPOS~\cite{Pierog:2013ria} and \Peight v208~\cite{Sjostrand:2007gs} generators
(with the CMS underlying event tune CUETP8M1~\cite{GEN-14-001}) indicate that 94--96\% of the
events satisfy the analysis selections,
i.e., they have at least one stable particle from the \pp interaction with energy $E>3$\GeV in each of the $\eta$
regions $-5<\eta <-3$ and $3<\eta <5$.

The present analysis is based on a sample of events with high-purity primary tracks \cite{TRK-11-001} originating from the
\pp interaction. To obtain this sample, additional requirements are applied.
The significance of the distance between the track and the primary vertex along the beam axis, $d_{z}/\sigma_{d_{z}}$,
and the significance of the impact parameter relative to the best resolution of the vertex
coordinates transverse to the beam, $d_{\mathrm{T}}/\sigma_{d_{\mathrm{T}}}$,
must both be less than 3 in absolute value, and the relative \pt uncertainty, $\sigma(\pt)/\pt$, must be less than 10\%.
To ensure high tracking efficiency and to reduce the rate of misreconstructed
tracks, primary tracks with $\abs{\eta}<2.4$ and $\pt > 0.1\GeVc$ are used in the analysis
(a \pt cutoff of 0.4\GeVc is used in the track multiplicity determination to match the HLT requirement).
Simulation studies based on \Peight
are used to obtain the geometrical acceptance and efficiency for primary track
reconstruction as well as the rate of misreconstructed tracks. The combined acceptance and efficiency is better than
60\% for $\pt>0.4$\GeVc and $\abs{\eta}<2.4$ and better than 90\% in the $\abs{\eta}<1$ region for $\pt>0.6\GeVc$. For the
track multiplicity range studied in this paper, no dependence of the tracking efficiency on track multiplicity
is found and the rate of misreconstructed tracks is 1--2\% according to simulations.

Following the procedure established in Refs.~\cite{Chatrchyan:2011eka,Chatrchyan:2012wg,CMS:2012qk,Chatrchyan:2013nka,Khachatryan:2014jra}, the data set is divided into classes of events with different track multiplicity, \noff,
which is evaluated by counting primary tracks with $\abs{\eta}<2.4$ and $\pt >0.4$\GeVc.
Details of the multiplicity classification in this analysis are provided
in Table~\ref{tab:multbinning}, which also gives the fraction with respect to the full
multiplicity distribution and the average number of primary tracks
before and after correcting for detector effects. The minimum bias
sample is used for the range $\noff < 80$, while various high-multiplicity samples are used for
\noff ranges above 80.

\begin{table}[ht]
\centering
\topcaption{\label{tab:multbinning}
Multiplicity classes used in the analysis, corresponding fraction of the
full event sample, observed and corrected average charged particle
multiplicities for $\abs{\eta}<2.4$ and $\pt >0.4$\GeVc. Systematic
uncertainties are given for the corrected multiplicities.
}
\newcolumntype{x}{D{,}{\,\pm\,}{-1}}
\newcolumntype{.}{D{.}{.}{-1}}
\newcolumntype{b}{D{,}{}{-1}}
\begin{scotch}{ l  .  b  x}
Multiplicity class (\noff) & \multicolumn{1}{c}{Fraction} & \multicolumn{1}{c}{$\left<\noff \right>$} & \multicolumn{1}{c}{$\left<N_\text{trk}^\text{corrected}\right>$}\\
\hline
Minimum bias           &   1.0  &  20,   & 23,1\\
$[2, 34]$ & 0.82 & 13, & 16,1 \\
$[35, 79]$ & 0.15 & 47, & 58,2 \\
$[80, 104]$  & 0.02 & 88, & 107,4 \\
$[105, 134]$ & \multicolumn{1}{c}{3.3$\times10^{-4}$} & 113, & 131,5  \\
${\ge}135$ & \multicolumn{1}{c}{1.4$\times10^{-5}$} & 145, & 168,7 \\
\end{scotch}
\end{table}

For each track multiplicity class, ``trigger'' particles are defined as charged particles originating
from the primary vertex within a given \pt range.
The number of trigger particles for each \pt range in the event is denoted by $N_\text{trig}$.
In this analysis, particle pairs are formed by associating every trigger particle with
the remaining charged primary particles (associated particles) from the same \pt interval
as the trigger particle. The per-trigger-particle associated yield is defined as
\begin{equation}
\label{2pcorr_incl}
\frac{1}{N_\text{trig}}\frac{\rd^{2}N^\text{pair}}{\rd\Delta\eta\, \rd\Delta\phi}
= B(0,0)\,\frac{S(\Delta\eta,\Delta\phi)}{B(\Delta\eta,\Delta\phi)},
\end{equation}
where $\Delta\eta$ and $\Delta\phi$ are the differences in $\eta$
and $\phi$ of the pair. The symbol $N^\text{pair}$ denotes the number of particle pairs.
The signal distribution, $S(\Delta\eta,\Delta\phi)$, is the per-trigger-particle yield of
particle pairs from the same event,
\begin{equation}
\label{eq:signal}
S(\Delta\eta,\Delta\phi) = \frac{1}{N_\text{trig}}\frac{\rd^{2}N^\text{same}}{\rd\Delta\eta \rd\Delta\phi}.
\end{equation}
The symbol $N^\text{same}$ denotes the number of pairs
taken from the same event. The mixed-event background distribution, used to account
for random combinatorial background and pair acceptance effects,
\begin{equation}
\label{eq:background}
B(\Delta\eta,\Delta\phi) = \frac{1}{N_\text{trig}}\frac{\rd^{2}N^\text{mix}}{\rd\Delta\eta\, \rd\Delta\phi},
\end{equation}
is constructed by pairing the trigger particles in each event with the
particles from 10 different random events within a 0.2\unit{cm} wide $z_\text{vtx}$ range.
The symbol $N^\text{mix}$ denotes the number of pairs taken from the mixed event,
while $B(0,0)$ represents the mixed-event associated yield for
both particles of the pair going in approximately the same direction and
thus having full pair acceptance (with a bin width of 0.3 in $\Delta\eta$ and $\pi/16$ in $\Delta\phi$).
Therefore, the ratio $B(0,0)/B(\Delta\eta,\Delta\phi)$
is the pair-acceptance correction factor used to derive the corrected
per-trigger-particle associated yield distribution. The signal and background distributions
are first calculated for each event, and then averaged over all the events within
the track multiplicity class for each \pt range.

Each reconstructed track is weighted by the inverse of an efficiency factor, which accounts
for the detector acceptance, the reconstruction efficiency, and the fraction of misreconstructed
tracks (the same factor as used for correcting the average multiplicity in Table~\ref{tab:multbinning}).

\begin{figure}[hbt]
  \centering
    \includegraphics[width=0.48\textwidth]{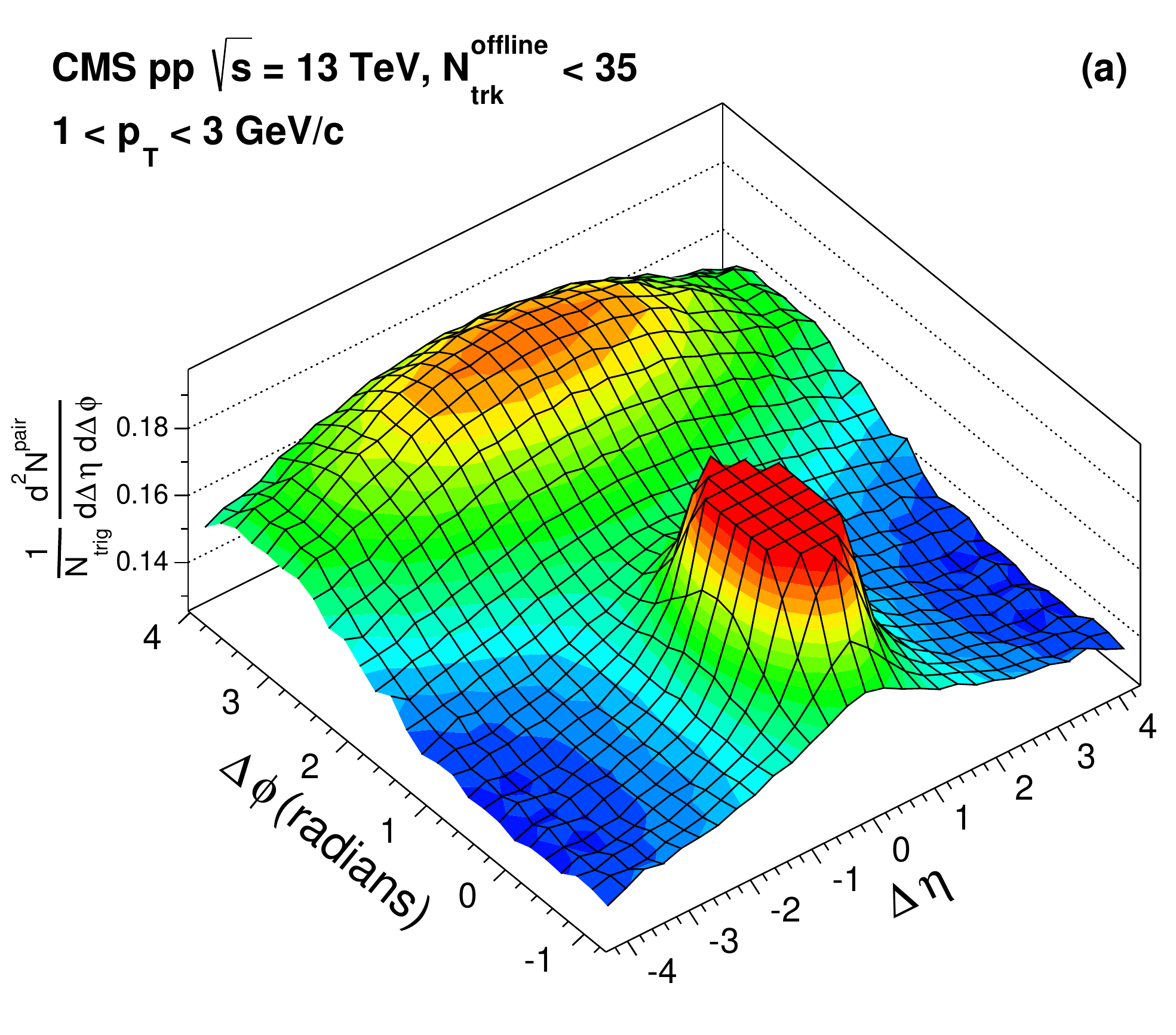}
    \includegraphics[width=0.48\textwidth]{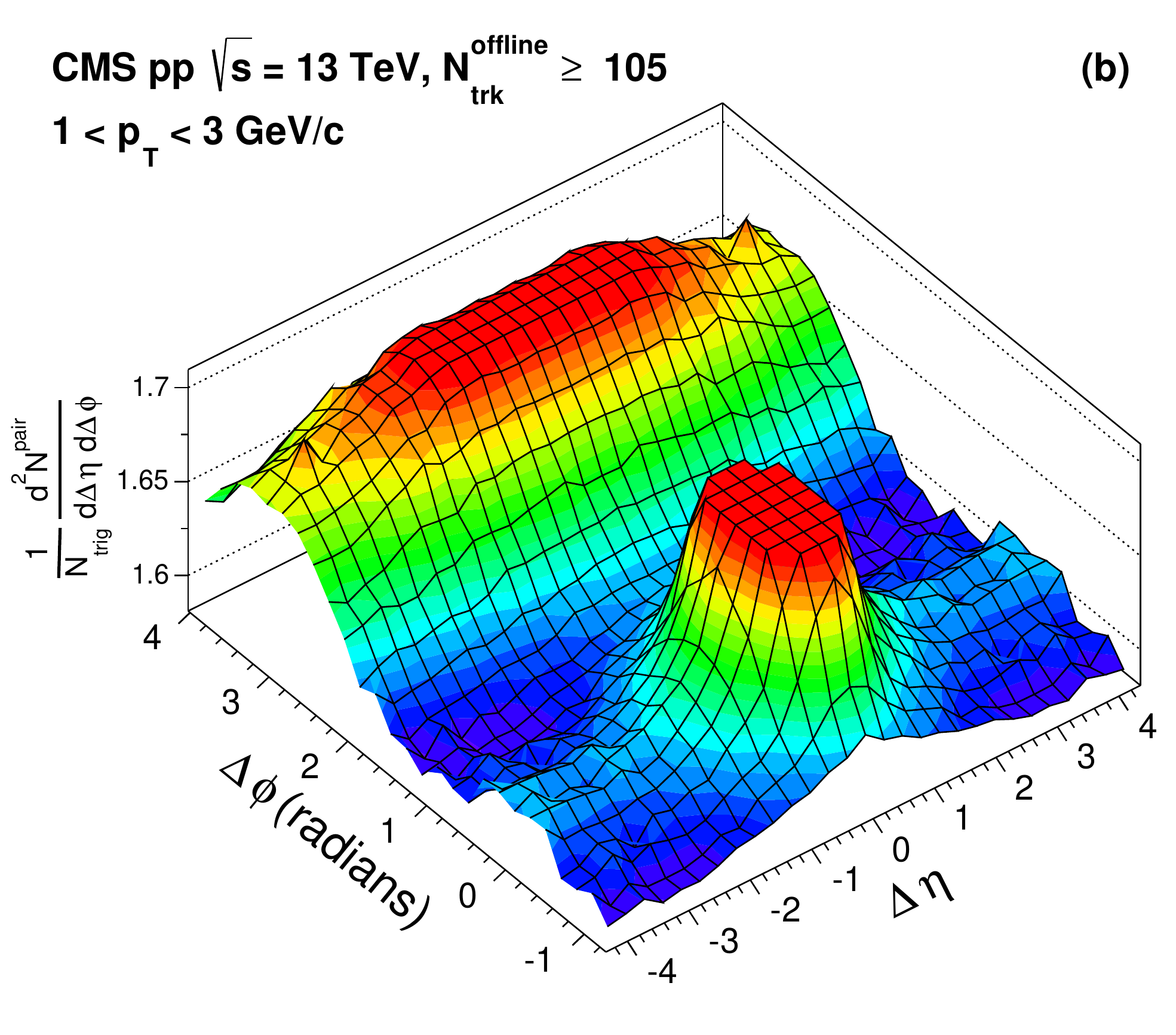}
    \caption{ The 2D $(\deta, \dphi)$ two-particle correlation functions in \pp collisions at $\roots = 13$\TeV     for pairs of charged particles both in the range $1<\pt<3$\GeVc. Results are shown for (a)~low-multiplicity
    events ($\noff < 35$) and for (b)~a high-multiplicity sample ($\noff \geq 105$).
    The sharp peaks from jet correlations around $(\deta, \dphi) = (0, 0)$ are truncated to better
    illustrate the long-range correlations.
    }
    \label{fig:ridge2D}
  \end{figure}

The two-dimensional (2D) \deta--\dphi\ two-particle correlation functions for events with low and high multiplicities
are shown in Fig.~\ref{fig:ridge2D}.
As in our earlier papers, pairs of charged particles both in the
range $1<\pt<3$\GeVc are used in this analysis.
For the low-multiplicity sample ($\noff < 35$), the dominant features are
the peak near $(\deta, \dphi) = (0, 0)$ (truncated for better illustration
of the long-range structures) for pairs of particles originating from the same jet. The
elongated structure at $\dphi \approx \pi$ corresponds to pairs of particles from back-to-back jets. In high-multiplicity \pp
events ($\noff \geq 105$), in addition to these jet-like correlation structures, a ``ridge''-like structure is clearly
visible at $\dphi \approx 0$, extending over a range of at least 4 units in $\abs{\deta}$.
No such long-range correlations are predicted by \PYTHIA.

\begin{figure*}[hbt]
  \centering
    \includegraphics[width=\textwidth]{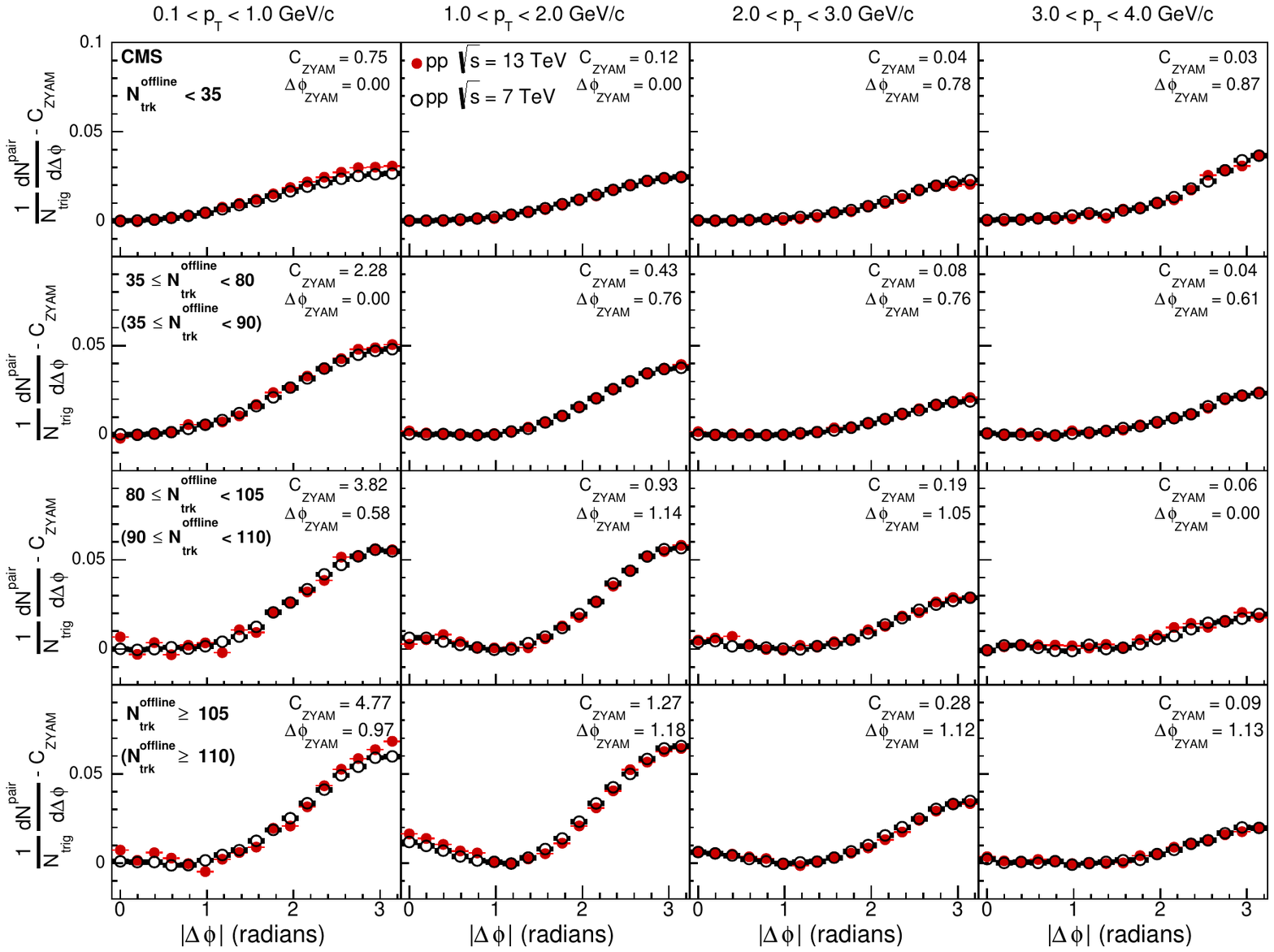}
    \caption{
Correlated yield obtained with the ZYAM procedure as a function of $\abs{\dphi}$, averaged over $2<\abs{\deta}<4$ in different
\pt and multiplicity bins for \pp data at $\roots = 13$\TeV (filled circles) and 7\TeV (open circles).
The \pt selection applies to both particles in the pair. Numbers in brackets indicate the multiplicity range of the 7\TeV data when different from that at 13\TeV. The statistical uncertainties are smaller than the marker size.
The subtracted ZYAM constant is given in each panel ($C_\mathrm{ZYAM}$).
    }
    \label{fig:ridge1D}
  \end{figure*}

To quantitatively investigate these long-range near-side correlations,
and to provide a direct comparison to \pp results at lower collision energy,
one-dimensional (1D) distributions in $\Delta\phi$ are constructed
by averaging the signal and background 2D distributions over
$2 < \abs{\deta} < 4$, as done in Refs.~\cite{Khachatryan:2010gv,CMS:2012qk,Chatrchyan:2013nka}.
The correlated portion of the associated yield is estimated by using an implementation of the
zero-yield-at-minimum (ZYAM) procedure~\cite{PhysRevC.72.011902}.
The 1D $\Delta\phi$ correlation function is fitted with a truncated Fourier series up to the fifth term.
The minimum value of the fit function, $C_\mathrm{ZYAM}$,
is then subtracted from the 1D \dphi\ correlation function
as a constant background (containing no information about correlations) so that the minimum of the correlation function is zero.
The location of the minimum of the function in this region is denoted as $\Delta\phi_\mathrm{ZYAM}$.
The ZYAM procedure is a straightforward way to quantify the magnitude of
long-range near-side yield. However, it does not take into account
potential biases introduced by away-side jet correlations leading to a
non-flat distribution on the near-side. Therefore, when performing
data-theory comparisons, other sources of correlations, such as jets,
should be included in the model calculation.

Figure~\ref{fig:ridge1D} shows the resulting $\Delta\phi$ correlation functions for various selections
in \pt and multiplicity \noff.
The results for \pp data at $\roots = 7$\TeV\ are also shown for comparison.
The selected \noff ranges in the 7 and 13\TeV data do not match precisely because of slight differences
in the multiplicity domains for which the high-multiplicity triggers used in 2010 and 2015 are fully efficient.
Note that the previously published \pp data at $\roots = 7$\TeV\ in Ref.~\cite{Khachatryan:2010gv} 
are obtained by means of a slightly different definition of the two-particle correlation functions 
and the 7\TeV\ data shown in Fig.~\ref{fig:ridge1D} have therefore been reanalysed. 
The difference has no impact on the associated yields for high-multiplicity events, and is 
only noticeable at very low multiplicity and high \pt, where most of the particle pairs are 
localised around (\deta,\dphi) $\sim$ (0,0) due to jet-like correlations.

Nearly no center-of-mass energy dependence is observed for the correlations 
in any \pt\ or multiplicity range, as shown in Fig.~\ref{fig:ridge1D}.
A clear evolution of the $\dphi$ correlation function with both \pt and \noff is observed at both collision energies. 
For the lowest multiplicity sample, the correlation
functions have a minimum at $\dphi = 0$ and a maximum at $\dphi = \pi$, reflecting the correlations from
momentum conservation and the increasing contribution from back-to-back jet-like correlations at higher \pt.
For high-multiplicity \pp events ($\noff \geqslant 80$), a second local maximum near $\abs{\dphi} \approx 0$
becomes visible, reflecting near-side, long-range correlations that appear as a ridge-like structure.
This near-side correlation signal is strongest in the $1 < \pt < 2$\GeVc range and increases with
multiplicity.

Based on the studies in Ref.~\cite{TRK-11-001}, the total systematic uncertainty of the tracking efficiency is 3.9\%,
which translates into a 3.9\% systematic uncertainty of the associated yields. The systematic uncertainties related to
the track quality requirements are studied by varying the track selections on $d_{z}/\sigma_{d_{z}}$
and $d_\mathrm{T}/\sigma_{d_\mathrm{T}}$
between 2 and 5. These changes produce effects on the associated yields smaller than 0.0006 in absolute value. In order
to evaluate the uncertainty of the trigger efficiency, results from high-multiplicity data collected with two different
triggers are compared. The results agree to better than 0.0015; this is taken as an estimate of the trigger efficiency
contribution to the systematic uncertainty. The possible contamination of residual pile-up events is investigated by comparing
the nominal results to those obtained without any pile-up rejection or with the requirement of only one reconstructed vertex.
The corresponding effect on the associated yield is less than 0.0006 in absolute value. The sensitivity of the results to the
vertex position along the beam direction ($z_\mathrm{vtx}$) is quantified by comparing results for $|z_\mathrm{vtx}|<3$\unit{cm} and
$3<\abs{z_\mathrm{vtx}}<15$\unit{cm}, which yields a contribution
to the systematic uncertainty of less than 0.0010. Finally, an alternative choice of a second-order polynomial fit function for estimating
$C_\mathrm{ZYAM}$ in the region $0.1<\abs{\Delta\phi}<2.0$ gives an absolute systematic uncertainty
of 0.0007 in the total correlated yield from the ZYAM procedure. The event multiplicity classification
is not varied in the systematic studies.
All the systematic effects studied yield contributions that are independent of
\pt and multiplicity; their values are summarized in Table~\ref{tab:syst-table-yield}.

\begin{table}[ht]
\topcaption{\label{tab:syst-table-yield} Summary of systematic uncertainties on the
long-range, near-side associated yields in \pp collisions at $\roots = 13$\TeV.}
\centering
\newcolumntype{.}{D{.}{.}{-1}}
\begin{scotch}{l.}
 Systematic uncertainty sources           & \multicolumn{1}{c}{Abs. uncertainty ($\times 10^{-3}$)} \\
\hline
 Track quality requirements              & 0.6 \\
 Trigger efficiency                      & 1.5 \\
 Correction for tracking efficiency      & \multicolumn{1}{c}{$<$0.08} \\
 Effect of pile-up events                & 0.6 \\
 Vertex selection                        & 1.0 \\
 ZYAM procedure                          & 0.7 \\
\hline
 Total                                   & 2.1  \\
\end{scotch}
\end{table}

\begin{figure*}[hbt]
  \centering
    \includegraphics[width=\textwidth]{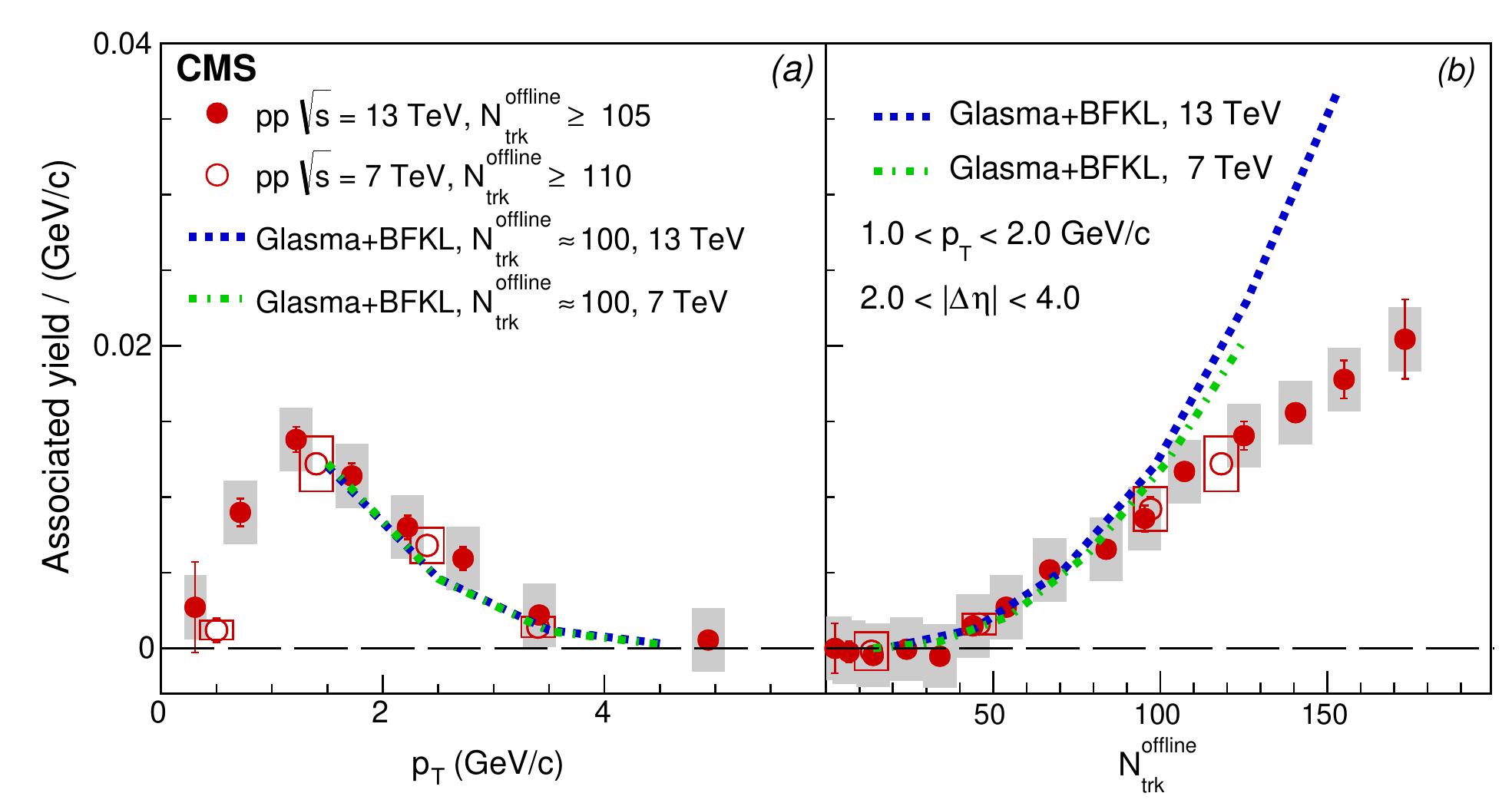}
    \caption{ Associated yield for the near-side
of the correlation function averaged over $2<\abs{\deta}<4$ and integrated over the region $\abs{\dphi} < \Delta\phi_\mathrm{ZYAM}$
for \pp data at $\roots = 13$\TeV (filled circles) and 7\TeV (open circles).
Panel (a) shows the associated yield as a function of \pt for events with $\noff \geq 105$. The \pt value for each \pt bin is
the average \pt value. In panel (b) the associated yield for $1 < \pt < 2$\GeVc is shown as a function
of the multiplicity, \noff.
The \noff value at which the yield is plotted is the average \noff value in the bin.
The \pt selection applies to both particles in each pair. The error bars correspond to the statistical uncertainties,
while the shaded areas and boxes denote the systematic uncertainties. Curves represent the predictions of the gluon saturation model~\cite{Dusling:2015rja}.
    }
    \label{fig:yield}
  \end{figure*}

The strength of the long-range, near-side correlations
can be further quantified by integrating the correlated yields from
Fig.~\ref{fig:ridge1D} over $\abs{\dphi} < \Delta\phi_\mathrm{ZYAM}$ for
each \pt range and event multiplicity class.
The resulting integrated near-side yield, divided by
the width of the \pt interval, is
plotted as a function of the particle \pt and the event multiplicity in Fig.~\ref{fig:yield}
for the present data. Finer \pt and \noff ranges than in Fig.~\ref{fig:ridge1D} are used for better
illustrating the trend of the data. The previous results from $\roots = 7$\TeV in wider
\pt and \noff ranges are also shown for comparison.
The 7\TeV data obtained from Ref.~\cite{CMS:2012qk}
are multiplied by two, as their range in \dphi\ is 0--$\Delta\phi_\mathrm{ZYAM}$,
half of the full near-side structure range.

Figure~\ref{fig:yield} (a) shows that the associated
yield of long-range near-side correlations for events with $\noff \geq 105$ ($\noff \geq 110$ for the 7\TeV data)
peaks in the region  $1 < \pt < 2$\GeVc for both center-of-mass energies.
The yield reaches a maximum around $\pt \approx 1$\GeVc and decreases with increasing \pt.
No center-of-mass energy dependence is visible. The multiplicity dependence of the associated yield for $1 < \pt < 2$\GeVc particle pairs
is shown in Fig.~\ref{fig:yield}~(b). For low-multiplicity events,
the associated yield determined with the ZYAM procedure
is consistent with zero. This indicates that ridge-like correlations are absent or smaller than
the negative correlations expected because of, for example, momentum conservation. At higher multiplicity the
ridge-like correlation emerges, with an approximately linear
rise of the associated yield with multiplicity for $\noff \gtrsim 40$.

In the framework of gluon saturation models, a long-range correlation structure is predicted
to arise from initial collimated gluon emissions~\cite{Dumitru:2010iy,Dusling:2012iga,Dusling:2013qoz}.
The energy dependence of associated yields observed in the data is qualitatively in agreement 
with this model at $\roots= 13$\TeV~\cite{Dusling:2015rja}, as shown in
Fig.~\ref{fig:yield} (b). However, although the model calculation quantitatively describes the associated yields
over the multiplicity range covered by the previous 7\TeV\ data, significant deviations are observed at the higher 
multiplicities probed by the present 13\TeV\ data. 
The associated yields predicted by this model exhibit a much
faster increase with \noff\ than that seen in the data, suggesting that other other mechanisms may be 
active in this region. Hydrodynamic models also predict no energy 
dependence: they reproduce the collective flow effect in heavy-ion 
collisions, which is nearly unchanged from the RHIC to the LHC 
center-of-mass energies, although they differ by more than an order of 
magnitude~\cite{Aamodt:2010pa,Chatrchyan:2012ta,ATLAS:2011ah}. However, it remains to be seen whether hydrodynamic 
models can quantitatively describe the behavior of the observables 
presented here.

\begin{figure}[hbt]
  \centering
    \includegraphics[width=\cmsFigWidth]{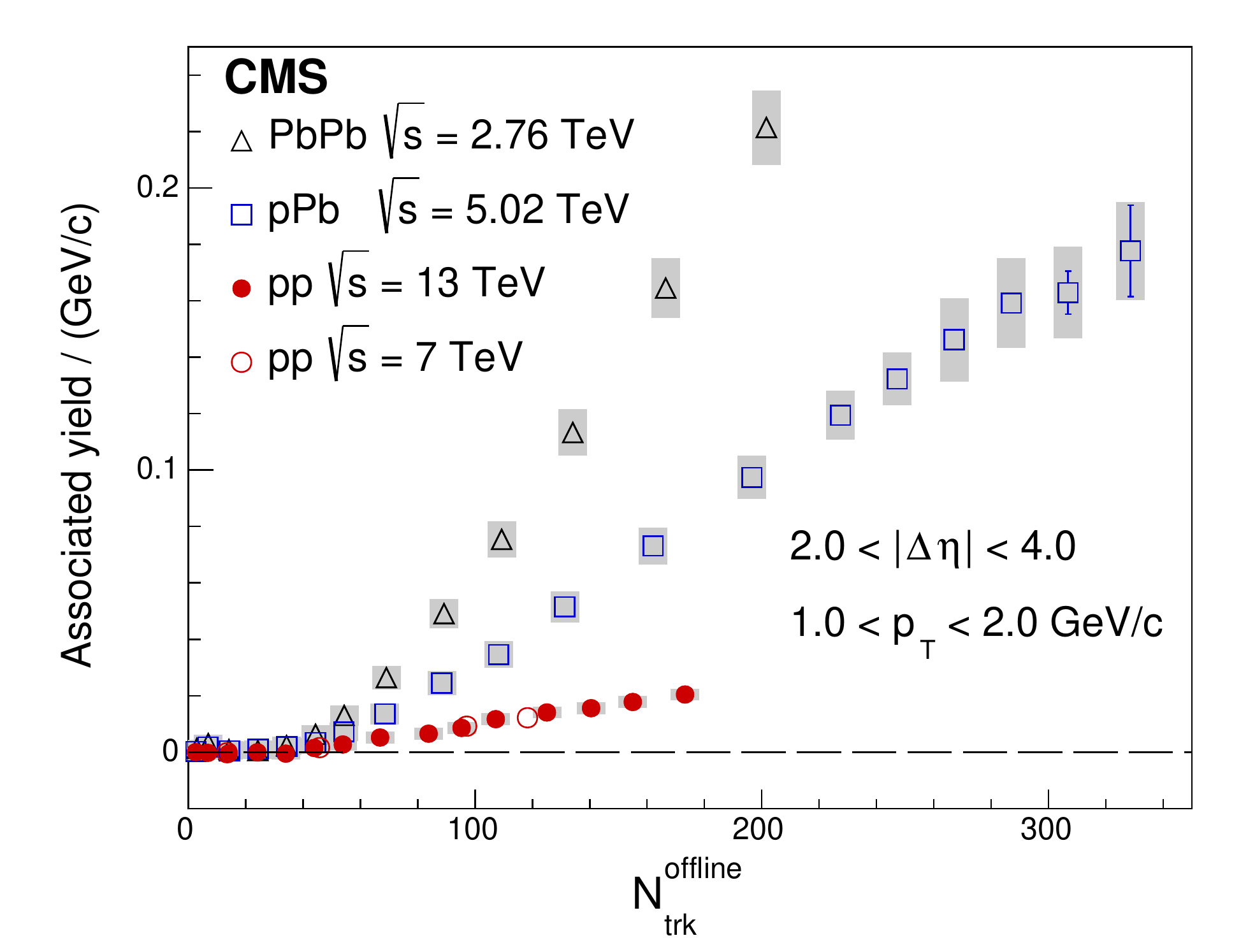}
    \caption{ Associated yield of long-range near-side two-particle correlations for $1 < \pt < 2$\GeVc
in \pp collisions at $\roots = 13$ and 7\TeV, \pPb collisions at $\rootsNN = 5.02$\TeV,
and \PbPb collisions at $\rootsNN = 2.76$\TeV. Associated yield for the near-side
of the correlation function is averaged over $2<\abs{\deta}<4$ and integrated over the region $\abs{\dphi} < \Delta\phi_\mathrm{ZYAM}$.
The \noff value for each \noff bin is
the average \noff value. The error bars correspond to the statistical uncertainties,
while the shaded areas denote the systematic uncertainties. Note that there are \PbPb points above the
upper vertical scale, which are not shown for clarity.
    }
    \label{fig:v2allsystems}
  \end{figure}

Long-range near-side yields have also been measured for \pPb and \PbPb collisions
by CMS~\cite{Chatrchyan:2013nka}.
Figure~\ref{fig:v2allsystems} compares the associated yields
in \pp, \pPb, and \PbPb collisions for $1 < \pt < 2$\GeVc
as a function of the track multiplicity. The various data sets were
collected at different center-of-mass energies, but this should have
negligible effect on the results, as discussed above. In all three systems, the ridge-like correlations become
significant at a multiplicity value of about 40, and exhibit a nearly linear increase for higher values.
For a given track multiplicity, the associated yield in \pp collisions is roughly 10\% and 25\% of
those observed in \PbPb and \pPb collisions, respectively. Clearly,
there is a strong collision system size dependence of the long-range near-side
correlations.

In summary, two-particle angular correlations in \pp collisions 
at $\roots = 13$\TeV have been measured by the CMS experiment at the LHC. The data
correspond to an integrated luminosity of about 270\nbinv.
As first observed in \pp\ collisions at $\roots = 7$\TeV,
two-particle azimuthal correlations in high-multiplicity \pp collisions
exhibit a long-range structure in the near-side ($\dphi \approx 0$)
extending over at least 4 units in pseudorapidity separation.
The effect is most evident in the intermediate transverse momentum region between 1 and 2\GeVc.
The near-side long-range yield obtained with the ZYAM procedure is found to be consistent
with zero in the low-multiplicity region, with an approximately linear increase
with multiplicity for $\noff \gtrsim 40$.
The new 13\TeV\ data presented in this Letter significantly 
extends the multiplicity coverage achieved by previously data at $\roots = 7$\TeV.
Finally, a strong collision system size dependence
is observed when comparing data from \pp, \pPb, and \PbPb collisions.
Comparing the \pp data at $\roots = 7$\TeV and 13\TeV,
no collision energy dependence of the near-side associated yields is
observed.

\begin{acknowledgments}
We congratulate our colleagues in the CERN accelerator departments for the excellent performance of the LHC and thank the technical and administrative staffs at CERN and at other CMS institutes for their contributions to the success of the CMS effort. In addition, we gratefully acknowledge the computing centers and personnel of the Worldwide LHC Computing Grid for delivering so effectively the computing infrastructure essential to our analyses. Finally, we acknowledge the enduring support for the construction and operation of the LHC and the CMS detector provided by the following funding agencies: BMWFW and FWF (Austria); FNRS and FWO (Belgium); CNPq, CAPES, FAPERJ, and FAPESP (Brazil); MES (Bulgaria); CERN; CAS, MoST, and NSFC (China); COLCIENCIAS (Colombia); MSES and CSF (Croatia); RPF (Cyprus); MoER, ERC IUT and ERDF (Estonia); Academy of Finland, MEC, and HIP (Finland); CEA and CNRS/IN2P3 (France); BMBF, DFG, and HGF (Germany); GSRT (Greece); OTKA and NIH (Hungary); DAE and DST (India); IPM (Iran); SFI (Ireland); INFN (Italy); MSIP and NRF (Republic of Korea); LAS (Lithuania); MOE and UM (Malaysia); CINVESTAV, CONACYT, SEP, and UASLP-FAI (Mexico); MBIE (New Zealand); PAEC (Pakistan); MSHE and NSC (Poland); FCT (Portugal); JINR (Dubna); MON, RosAtom, RAS and RFBR (Russia); MESTD (Serbia); SEIDI and CPAN (Spain); Swiss Funding Agencies (Switzerland); MST (Taipei); ThEPCenter, IPST, STAR and NSTDA (Thailand); TUBITAK and TAEK (Turkey); NASU and SFFR (Ukraine); STFC (United Kingdom); DOE and NSF (USA).
\end{acknowledgments}

\bibliography{auto_generated}

\providecommand{\href}[2]{#2}\begingroup\raggedright\begin{thebibliography}{10}%
\makeatletter
\providecommand{\hrefCMSnoop }[0]{\@secondoftwo}%
\makeatother
\providecommand{\doi}{\texttt{doi:}\begingroup \urlstyle{tt}\Url}

\bibitem{Alver:2008gk}
\hrefCMSnoop {}{{PHOBOS} Collaboration, ``{System size dependence of cluster
  properties from two- particle angular correlations in Cu+Cu and Au+Au
  collisions at \rootsNN\ = 200\GeV}'',} \textit{ Phys. Rev. C} \textbf{ 81}
  (2010) 024904,
  \href{http://dx.doi.org/10.1103/PhysRevC.81.024904}{\doi{10.1103/PhysRevC.81.024904}},
\href{http://www.arXiv.org/abs/0812.1172}{\texttt{arXiv:0812.1172}}.

\bibitem{Alver:2009id}
\hrefCMSnoop {}{{PHOBOS} Collaboration, ``{High transverse momentum triggered
  correlations over a large pseudorapidity acceptance in Au+Au collisions at
  \rootsNN\ = 200\GeV}'',} \textit{ Phys. Rev. Lett.} \textbf{ 104} (2010)
  062301,
  \href{http://dx.doi.org/10.1103/PhysRevLett.104.062301}{\doi{10.1103/PhysRevLett.104.062301}},
\href{http://www.arXiv.org/abs/0903.2811}{\texttt{arXiv:0903.2811}}.

\bibitem{Abelev:2009jv}
\hrefCMSnoop {}{{STAR} Collaboration, ``{Three-particle coincidence of the long
  range pseudorapidity correlation in high energy nucleus-nucleus
  collisions}'',} \textit{ Phys. Rev. Lett.} \textbf{ 105} (2010) 022301,
  \href{http://dx.doi.org/10.1103/PhysRevLett.105.022301}{\doi{10.1103/PhysRevLett.105.022301}},
\href{http://www.arXiv.org/abs/0912.3977}{\texttt{arXiv:0912.3977}}.

\bibitem{Alver:2010gr}
\hrefCMSnoop {}{B.~Alver and G.~Roland, ``{Collision geometry fluctuations and
  triangular flow in heavy-ion collisions}'',} \textit{ Phys. Rev. C} \textbf{
  81} (2010) 054905,
  \href{http://dx.doi.org/10.1103/PhysRevC.81.054905}{\doi{10.1103/PhysRevC.81.054905}},
\href{http://www.arXiv.org/abs/1003.0194}{\texttt{arXiv:1003.0194}}.

\bibitem{Alver:2010dn}
\hrefCMSnoop {}{B.~H. Alver, C.~Gombeaud, M.~Luzum, and J.-Y. Ollitrault,
  ``{Triangular flow in hydrodynamics and transport theory}'',} \textit{ Phys.
  Rev. C} \textbf{ 82} (2010) 034913,
  \href{http://dx.doi.org/10.1103/PhysRevC.82.034913}{\doi{10.1103/PhysRevC.82.034913}},
\href{http://www.arXiv.org/abs/1007.5469}{\texttt{arXiv:1007.5469}}.

\bibitem{Schenke:2010rr}
\hrefCMSnoop {}{B.~Schenke, S.~Jeon, and C.~Gale, ``Elliptic and triangular
  flow in event-by-event {D=3+1} viscous hydrodynamics'',} \textit{ Phys. Rev.
  Lett.} \textbf{ 106} (2011) 042301,
  \href{http://dx.doi.org/10.1103/PhysRevLett.106.042301}{\doi{10.1103/PhysRevLett.106.042301}},
\href{http://www.arXiv.org/abs/1009.3244}{\texttt{arXiv:1009.3244}}.

\bibitem{Petersen:2010cw}
\hrefCMSnoop {}{H.~Petersen, G.-Y. Qin, S.~A. Bass, and B.~M{\"u}ller,
  ``{Triangular flow in event-by-event ideal hydrodynamics in Au+Au collisions
  at \rootsNN\ = 200A GeV}'',} \textit{ Phys. Rev. C} \textbf{ 82} (2010)
  041901,
  \href{http://dx.doi.org/10.1103/PhysRevC.82.041901}{\doi{10.1103/PhysRevC.82.041901}},
\href{http://www.arXiv.org/abs/1008.0625}{\texttt{arXiv:1008.0625}}.

\bibitem{Xu:2010du}
\hrefCMSnoop {}{J.~Xu and C.~M. Ko, ``Effects of triangular flow on dihadron
  azimuthal correlations in relativistic heavy ion collisions'',} \textit{
  Phys. Rev. C} \textbf{ 83} (2011) 021903,
  \href{http://dx.doi.org/10.1103/PhysRevC.83.021903}{\doi{10.1103/PhysRevC.83.021903}},
\href{http://www.arXiv.org/abs/1011.3750}{\texttt{arXiv:1011.3750}}.

\bibitem{Teaney:2010vd}
\hrefCMSnoop {}{D.~Teaney and L.~Yan, ``Triangularity and dipole asymmetry in
  heavy ion collisions'',} \textit{ Phys. Rev. C} \textbf{ 83} (2011) 064904,
  \href{http://dx.doi.org/10.1103/PhysRevC.83.064904}{\doi{10.1103/PhysRevC.83.064904}},
\href{http://www.arXiv.org/abs/1010.1876}{\texttt{arXiv:1010.1876}}.

\bibitem{Khachatryan:2010gv}
\hrefCMSnoop {}{{CMS Collaboration}, ``{Observation of long-range near-side
  angular correlations in proton-proton collisions at the LHC}'',} \textit{
  JHEP} \textbf{ 09} (2010) 091,
  \href{http://dx.doi.org/10.1007/JHEP09(2010)091}{\doi{10.1007/JHEP09(2010)091}},
\href{http://www.arXiv.org/abs/1009.4122}{\texttt{arXiv:1009.4122}}.

\bibitem{CMS:2012qk}
\hrefCMSnoop {}{{CMS Collaboration}, ``{Observation of long-range near-side
  angular correlations in pPb collisions at the LHC}'',} \textit{ Phys. Lett.
  B} \textbf{ 718} (2013) 795,
  \href{http://dx.doi.org/10.1016/j.physletb.2012.11.025}{\doi{10.1016/j.physletb.2012.11.025}},
\href{http://www.arXiv.org/abs/1210.5482}{\texttt{arXiv:1210.5482}}.

\bibitem{Abelev:2012ola}
\hrefCMSnoop {}{{ALICE Collaboration}, ``{Long-range angular correlations on
  the near and away side in pPb collisions at \rootsNN\ = 5.02\TeV}'',}
  \textit{ Phys. Lett. B} \textbf{ 719} (2013) 29,
  \href{http://dx.doi.org/10.1016/j.physletb.2013.01.012}{\doi{10.1016/j.physletb.2013.01.012}},
\href{http://www.arXiv.org/abs/1212.2001}{\texttt{arXiv:1212.2001}}.

\bibitem{Aad:2012gla}
\hrefCMSnoop {}{{ATLAS Collaboration}, ``{Observation of Associated Near-Side
  and Away-Side Long-Range Correlations in \rootsNN\ = 5.02\TeV Proton-Lead
  Collisions with the ATLAS Detector}'',} \textit{ Phys. Rev. Lett.} \textbf{
  110} (2013) 182302,
  \href{http://dx.doi.org/10.1103/PhysRevLett.110.182302}{\doi{10.1103/PhysRevLett.110.182302}},
\href{http://www.arXiv.org/abs/1212.5198}{\texttt{arXiv:1212.5198}}.

\bibitem{Chatrchyan:2013nka}
\hrefCMSnoop {}{{CMS} Collaboration, ``{Multiplicity and transverse momentum
  dependence of two- and four-particle correlations in pPb and \PbPb\
  collisions}'',} \textit{ Phys. Lett. B} \textbf{ 724} (2013) 213,
  \href{http://dx.doi.org/10.1016/j.physletb.2013.06.028}{\doi{10.1016/j.physletb.2013.06.028}},
\href{http://www.arXiv.org/abs/1305.0609}{\texttt{arXiv:1305.0609}}.

\bibitem{Khachatryan:2014jra}
\hrefCMSnoop {}{{CMS Collaboration}, ``{Long-range two-particle correlations of
  strange hadrons with charged particles in pPb and PbPb collisions at LHC
  energies}'',} \textit{ Phys. Lett. B} \textbf{ 742} (2015) 200,
  \href{http://dx.doi.org/10.1016/j.physletb.2015.01.034}{\doi{10.1016/j.physletb.2015.01.034}},
\href{http://www.arXiv.org/abs/1409.3392}{\texttt{arXiv:1409.3392}}.

\bibitem{Khachatryan:2015waa}
\hrefCMSnoop {}{{CMS Collaboration}, ``{Evidence for Collective Multiparticle
  Correlations in $p$-Pb Collisions}'',} \textit{ Phys. Rev. Lett.} \textbf{
  115} (2015) 012301,
  \href{http://dx.doi.org/10.1103/PhysRevLett.115.012301}{\doi{10.1103/PhysRevLett.115.012301}},
\href{http://www.arXiv.org/abs/1502.05382}{\texttt{arXiv:1502.05382}}.

\bibitem{Khachatryan:2015oea}
\hrefCMSnoop {}{{CMS Collaboration}, ``{Evidence for transverse momentum and
  pseudorapidity dependent event plane fluctuations in PbPb and pPb
  collisions}'',} \textit{ Phys. Rev. C} \textbf{ 92} (2015) 034911,
  \href{http://dx.doi.org/10.1103/PhysRevC.92.034911}{\doi{10.1103/PhysRevC.92.034911}},
\href{http://www.arXiv.org/abs/1503.01692}{\texttt{arXiv:1503.01692}}.

\bibitem{Aad:2013fja}
\hrefCMSnoop {}{{ATLAS Collaboration}, ``{Measurement with the ATLAS detector
  of multi-particle azimuthal correlations in p+Pb collisions at \rootsNN\ =
  5.02\TeV}'',} \textit{ Phys. Lett. B} \textbf{ 725} (2013) 60,
  \href{http://dx.doi.org/10.1016/j.physletb.2013.06.057}{\doi{10.1016/j.physletb.2013.06.057}},
\href{http://www.arXiv.org/abs/1303.2084}{\texttt{arXiv:1303.2084}}.

\bibitem{Aad:2014lta}
\hrefCMSnoop {}{{ATLAS Collaboration}, ``{Measurement of long-range
  pseudorapidity correlations and azimuthal harmonics in \rootsNN\ = 5.02\TeV\
  proton-lead collisions with the ATLAS detector}'',} \textit{ Phys. Rev. C}
  \textbf{ 90} (2014) 044906,
  \href{http://dx.doi.org/10.1103/PhysRevC.90.044906}{\doi{10.1103/PhysRevC.90.044906}},
\href{http://www.arXiv.org/abs/1409.1792}{\texttt{arXiv:1409.1792}}.

\bibitem{Abelev:2014mda}
\hrefCMSnoop {}{{ALICE Collaboration}, ``{Multiparticle azimuthal correlations
  in p-Pb and Pb-Pb collisions at the CERN Large Hadron Collider}'',} \textit{
  Phys. Rev. C} \textbf{ 90} (2014) 054901,
  \href{http://dx.doi.org/10.1103/PhysRevC.90.054901}{\doi{10.1103/PhysRevC.90.054901}},
\href{http://www.arXiv.org/abs/1406.2474}{\texttt{arXiv:1406.2474}}.

\bibitem{ABELEV:2013wsa}
\hrefCMSnoop {}{{ALICE Collaboration}, ``{Long-range angular correlations of
  $\pi$, K and p in p--Pb collisions at \rootsNN\ = 5.02\TeV}'',} \textit{
  Phys. Lett. B} \textbf{ 726} (2013) 164,
  \href{http://dx.doi.org/10.1016/j.physletb.2013.08.024}{\doi{10.1016/j.physletb.2013.08.024}},
\href{http://www.arXiv.org/abs/1307.3237}{\texttt{arXiv:1307.3237}}.

\bibitem{Li:2012hc}
\hrefCMSnoop {}{W.~Li, ``{Observation of a ridge correlation structure in high
  multiplicity proton-proton collisions: A brief review}'',} \textit{ Mod.
  Phys. Lett. A} \textbf{ 27} (2012) 1230018,
  \href{http://dx.doi.org/10.1142/S0217732312300182}{\doi{10.1142/S0217732312300182}},
\href{http://www.arXiv.org/abs/1206.0148}{\texttt{arXiv:1206.0148}}.

\bibitem{Bozek:2011if}
\hrefCMSnoop {}{P.~Bo$\dot{\text{z}}$ek, ``{Collective flow in p-Pb and d-Pb
  collisions at TeV energies}'',} \textit{ Phys. Rev. C} \textbf{ 85} (2012)
  014911,
  \href{http://dx.doi.org/10.1103/PhysRevC.85.014911}{\doi{10.1103/PhysRevC.85.014911}},
\href{http://www.arXiv.org/abs/1112.0915}{\texttt{arXiv:1112.0915}}.

\bibitem{Bozek:2012gr}
\hrefCMSnoop {}{P.~Bo$\dot{\text{z}}$ek and W.~Broniowski, ``{Correlations from
  hydrodynamic flow in pPb collisions}'',} \textit{ Phys. Lett. B} \textbf{
  718} (2013) 1557,
  \href{http://dx.doi.org/10.1016/j.physletb.2012.12.051}{\doi{10.1016/j.physletb.2012.12.051}},
\href{http://www.arXiv.org/abs/1211.0845}{\texttt{arXiv:1211.0845}}.

\bibitem{Dusling:2012cg}
\hrefCMSnoop {}{K.~Dusling and R.~Venugopalan, ``Evidence for {BFKL} and
  saturation dynamics from dihadron spectra at the {LHC}'',} \textit{ Phys.
  Rev. D} \textbf{ 87} (2013) 051502,
  \href{http://dx.doi.org/10.1103/PhysRevD.87.051502}{\doi{10.1103/PhysRevD.87.051502}},
\href{http://www.arXiv.org/abs/1210.3890}{\texttt{arXiv:1210.3890}}.

\bibitem{Dumitru:2014vka}
\hrefCMSnoop {}{A.~Dumitru and V.~Skokov, ``{Anisotropy of the semiclassical
  gluon field of a large nucleus at high energy}'',} \textit{ Phys. Rev. D}
  \textbf{ 91} (2015) 074006,
  \href{http://dx.doi.org/10.1103/PhysRevD.91.074006}{\doi{10.1103/PhysRevD.91.074006}},
\href{http://www.arXiv.org/abs/1411.6630}{\texttt{arXiv:1411.6630}}.

\bibitem{Schenke:2015aqa}
\hrefCMSnoop {}{B.~Schenke, S.~Schlichting, and R.~Venugopalan, ``{Azimuthal
  anisotropies in p$+$Pb collisions from classical Yang--Mills dynamics}'',}
  \textit{ Phys. Lett. B} \textbf{ 747} (2015) 76,
  \href{http://dx.doi.org/10.1016/j.physletb.2015.05.051}{\doi{10.1016/j.physletb.2015.05.051}},
\href{http://www.arXiv.org/abs/1502.01331}{\texttt{arXiv:1502.01331}}.

\bibitem{Aad:2015gqa}
\hrefCMSnoop {}{{ATLAS Collaboration}, ``{Observation of long-range elliptic
  anisotropies in $\sqrt{s}=$13 and 2.76 TeV pp collisions with the ATLAS
  detector}'',} (2015).
\href{http://www.arXiv.org/abs/1509.04776}{\texttt{arXiv:1509.04776}}.

\bibitem{TRK-11-001}
\hrefCMSnoop {}{{CMS Collaboration}, ``{Description and performance of track
  and primary-vertex reconstruction with the CMS tracker}'',} \textit{ JINST}
  \textbf{ 9} (2014) P10009,
  \href{http://dx.doi.org/10.1088/1748-0221/9/10/P10009}{\doi{10.1088/1748-0221/9/10/P10009}},
\href{http://www.arXiv.org/abs/1405.6569}{\texttt{arXiv:1405.6569}}.

\bibitem{Chatrchyan:2008zzk}
\hrefCMSnoop {}{{CMS Collaboration}, ``The {CMS} experiment at the {CERN}
  {LHC}'',} \textit{ JINST} \textbf{ 3} (2008) S08004,
\href{http://dx.doi.org/10.1088/1748-0221/3/08/S08004}{\doi{10.1088/1748-0221/3/08/S08004}}.

\bibitem{GEANT4}
\hrefCMSnoop {}{{Geant4} Collaboration, ``{Geant4---a simulation toolkit}'',}
  \textit{ Nucl. Instrum. Meth. A} \textbf{ 506} (2003) 250,
\href{http://dx.doi.org/10.1016/S0168-9002(03)01368-8}{\doi{10.1016/S0168-9002(03)01368-8}}.

\bibitem{Khachatryan:2010xs}
\hrefCMSnoop {}{{CMS Collaboration}, ``{Transverse momentum and pseudorapidity
  distributions of charged hadrons in pp collisions at $\sqrt{s} = 0.9$ and
  2.36 TeV}'',} \textit{ JHEP} \textbf{ 02} (10) 041,
  \href{http://dx.doi.org/10.1007/JHEP02(2010)041}{\doi{10.1007/JHEP02(2010)041}},
\href{http://www.arXiv.org/abs/1002.0621}{\texttt{arXiv:1002.0621}}.

\bibitem{Pierog:2013ria}
T.~Pierog\hrefCMSnoop {}{ {et~al.}, ``EPOS LHC: Test of collective
  hadronization with data measured at the CERN Large Hadron Collider'',}
  \textit{ Phys. Rev. C} \textbf{ 92} (Sep, 2015) 034906,
  \href{http://dx.doi.org/10.1103/PhysRevC.92.034906}{\doi{10.1103/PhysRevC.92.034906}}.

\bibitem{Sjostrand:2007gs}
\hrefCMSnoop {}{T.~Sj{\"o}strand, S.~Mrenna, and P.~Skands, ``{A Brief
  Introduction to PYTHIA 8.1}'',} \textit{ Comput. Phys. Commun.} \textbf{ 178}
  (2008) 852,
  \href{http://dx.doi.org/10.1016/j.cpc.2008.01.036}{\doi{10.1016/j.cpc.2008.01.036}},
\href{http://www.arXiv.org/abs/0710.3820}{\texttt{arXiv:0710.3820}}.

\bibitem{GEN-14-001}
\href {http://cdsweb.cern.ch/record/1697700}{{CMS Collaboration}, ``Underlying
  Event Tunes and Double Parton Scattering'',} CMS Physics Analysis Summary
  CMS-PAS-GEN-14-001, 2014.

\bibitem{Chatrchyan:2011eka}
\hrefCMSnoop {}{{CMS Collaboration}, ``{Long-range and short-range dihadron
  angular correlations in central PbPb collisions at \rootsNN\ = 2.76\TeV}'',}
  \textit{ JHEP} \textbf{ 07} (2011) 076,
  \href{http://dx.doi.org/10.1007/JHEP07(2011)076}{\doi{10.1007/JHEP07(2011)076}},
\href{http://www.arXiv.org/abs/1105.2438}{\texttt{arXiv:1105.2438}}.

\bibitem{Chatrchyan:2012wg}
\hrefCMSnoop {}{{CMS Collaboration}, ``{Centrality dependence of dihadron
  correlations and azimuthal anisotropy harmonics in PbPb collisions at
  \rootsNN\ = 2.76 TeV}'',} \textit{ Eur. Phys. J. C} \textbf{ 72} (2012) 2012,
  \href{http://dx.doi.org/10.1140/epjc/s10052-012-2012-3}{\doi{10.1140/epjc/s10052-012-2012-3}},
\href{http://www.arXiv.org/abs/1201.3158}{\texttt{arXiv:1201.3158}}.

\bibitem{PhysRevC.72.011902}
N.~N. Ajitanand\hrefCMSnoop {}{ {et~al.}, ``Decomposition of harmonic and jet
  contributions to particle-pair correlations at ultra-relativistic
  energies'',} \textit{ Phys. Rev. C} \textbf{ 72} (2005) 011902,
  \href{http://dx.doi.org/10.1103/PhysRevC.72.011902}{\doi{10.1103/PhysRevC.72.011902}},
\href{http://www.arXiv.org/abs/nucl-ex/0501025}{\texttt{arXiv:nucl-ex/0501025}}.

\bibitem{Dusling:2015rja}
\hrefCMSnoop {}{K.~Dusling, P.~Tribedy, and R.~Venugopalan, ``{Energy
  dependence of the ridge in high multiplicity proton-proton collisions}'',}
  (2015).
  \href{http://www.arXiv.org/abs/1509.04410}{\texttt{arXiv:1509.04410}}.
Submitted to PRD.

\bibitem{Dumitru:2010iy}
A.~Dumitru\hrefCMSnoop {}{ {et~al.}, ``The ridge in proton-proton collisions at
  the {LHC}'',} \textit{ Phys. Lett. B} \textbf{ 697} (2011) 21,
  \href{http://dx.doi.org/10.1016/j.physletb.2011.01.024}{\doi{10.1016/j.physletb.2011.01.024}},
\href{http://www.arXiv.org/abs/1009.5295}{\texttt{arXiv:1009.5295}}.

\bibitem{Dusling:2012iga}
\hrefCMSnoop {}{K.~Dusling and R.~Venugopalan, ``{Azimuthal collimation of long
  range rapidity correlations by strong color fields in high multiplicity
  hadron-hadron collisions}'',} \textit{ Phys. Rev. Lett.} \textbf{ 108} (2012)
  262001,
  \href{http://dx.doi.org/10.1103/PhysRevLett.108.262001}{\doi{10.1103/PhysRevLett.108.262001}},
\href{http://www.arXiv.org/abs/1201.2658}{\texttt{arXiv:1201.2658}}.

\bibitem{Dusling:2013qoz}
\hrefCMSnoop {}{K.~Dusling and R.~Venugopalan, ``{Comparison of the color glass
  condensate to dihadron correlations in proton-proton and proton-nucleus
  collisions}'',} \textit{ Phys. Rev. D} \textbf{ 87} (2013) 094034,
  \href{http://dx.doi.org/10.1103/PhysRevD.87.094034}{\doi{10.1103/PhysRevD.87.094034}},
\href{http://www.arXiv.org/abs/1302.7018}{\texttt{arXiv:1302.7018}}.

\bibitem{Aamodt:2010pa}
\hrefCMSnoop {}{{ALICE Collaboration}, ``{Elliptic flow of charged particles in
  Pb-Pb collisions at 2.76 TeV}'',} \textit{ Phys. Rev. Lett.} \textbf{ 105}
  (2010) 252302,
  \href{http://dx.doi.org/10.1103/PhysRevLett.105.252302}{\doi{10.1103/PhysRevLett.105.252302}},
\href{http://www.arXiv.org/abs/1011.3914}{\texttt{arXiv:1011.3914}}.

\bibitem{Chatrchyan:2012ta}
\hrefCMSnoop {}{{CMS Collaboration}, ``{Measurement of the elliptic anisotropy
  of charged particles produced in PbPb collisions at nucleon-nucleon
  center-of-mass energy = 2.76 TeV}'',} \textit{ Phys. Rev. C} \textbf{ 87}
  (2013) 014902,
  \href{http://dx.doi.org/10.1103/PhysRevC.87.014902}{\doi{10.1103/PhysRevC.87.014902}},
\href{http://www.arXiv.org/abs/1204.1409}{\texttt{arXiv:1204.1409}}.

\bibitem{ATLAS:2011ah}
\hrefCMSnoop {}{{ATLAS} Collaboration, ``{Measurement of the pseudorapidity and
  transverse momentum dependence of the elliptic flow of charged particles in
  lead-lead collisions at \rootsNN\ = 2.76\TeV\ with the ATLAS detector}'',}
  \textit{ Phys. Lett. B} \textbf{ 707} (2012) 330,
  \href{http://dx.doi.org/10.1016/j.physletb.2011.12.056}{\doi{10.1016/j.physletb.2011.12.056}},
\href{http://www.arXiv.org/abs/1108.6018}{\texttt{arXiv:1108.6018}}.

\end{thebibliography}\endgroup
\cleardoublepage \appendix\section{The CMS Collaboration \label{app:collab}}\begin{sloppypar}\hyphenpenalty=5000\widowpenalty=500\clubpenalty=5000\textbf{Yerevan Physics Institute,  Yerevan,  Armenia}\\*[0pt]
V.~Khachatryan, A.M.~Sirunyan, A.~Tumasyan
\vskip\cmsinstskip
\textbf{Institut f\"{u}r Hochenergiephysik der OeAW,  Wien,  Austria}\\*[0pt]
W.~Adam, E.~Asilar, T.~Bergauer, J.~Brandstetter, E.~Brondolin, M.~Dragicevic, J.~Er\"{o}, M.~Flechl, M.~Friedl, R.~Fr\"{u}hwirth\cmsAuthorMark{1}, V.M.~Ghete, C.~Hartl, N.~H\"{o}rmann, J.~Hrubec, M.~Jeitler\cmsAuthorMark{1}, V.~Kn\"{u}nz, A.~K\"{o}nig, M.~Krammer\cmsAuthorMark{1}, I.~Kr\"{a}tschmer, D.~Liko, T.~Matsushita, I.~Mikulec, D.~Rabady\cmsAuthorMark{2}, B.~Rahbaran, H.~Rohringer, J.~Schieck\cmsAuthorMark{1}, R.~Sch\"{o}fbeck, J.~Strauss, W.~Treberer-Treberspurg, W.~Waltenberger, C.-E.~Wulz\cmsAuthorMark{1}
% ...
\vskip\cmsinstskip
% ...
1:~~Also at Vienna University of Technology, Vienna, Austria\\
2:~~Also at CERN, European Organization for Nuclear Research, Geneva, Switzerland\\
% ...

\end{sloppypar}
\end{document}